\documentclass[aps,prl,reprint,groupedaddress]{revtex4-1} 
\usepackage{graphicx}
\usepackage{amsmath,amssymb}
\usepackage{textcomp}
\begin{document}

\title{Phase separation and pair condensation in a spin-imbalanced 2D Fermi gas}

\author{Debayan Mitra}
\affiliation{Department of Physics, Princeton University, Princeton, New Jersey 08540, USA}
\author{Peter T. Brown}
\affiliation{Department of Physics, Princeton University, Princeton, New Jersey 08540, USA}
\author{Peter Schau{\ss}}
\affiliation{Department of Physics, Princeton University, Princeton, New Jersey 08540, USA}
\author{Stanimir S. Kondov}
\affiliation{Department of Physics, Princeton University, Princeton, New Jersey 08540, USA}
\author{Waseem S. Bakr}
\affiliation{Department of Physics, Princeton University, Princeton, New Jersey 08540, USA}

\date{\today}

\begin{abstract}
We study a two-component quasi-two-dimensional Fermi gas with imbalanced spin populations. We probe the gas at different interaction strengths and polarizations by measuring the density of each spin component in the trap and the pair momentum distribution after time of flight. For a wide range of experimental parameters, we observe in-trap phase separation characterized by the appearance of a spin-balanced core surrounded by a polarized gas. Our momentum space measurements indicate pair condensation in the imbalanced gas even for large polarizations where phase separation vanishes, pointing to the presence of a polarized pair condensate. Our observation of zero momentum pair condensates in 2D spin-imbalanced gases opens the way to explorations of more exotic superfluid phases that occupy a large part of the phase diagram in lower dimensions.
\end{abstract}


\maketitle


Fermionic superfluids described by standard Bardeen-Cooper-Schrieffer theory are momentum-space condensates of Cooper pairs of opposite spins. Imbalancing the chemical potentials of the two spin species disrupts the Cooper pairing mechanism and can give rise to many interesting scenarios. For a small difference in the chemical potentials, the Fermi gas remains a spin-balanced superfluid. As the chemical potential imbalance is increased, it eventually becomes comparable to the superfluid gap. At this point, known as the Clogston-Chandrasekhar limit \cite{Clogston1962}, the gas becomes polarized but superfluidity may persist due to the presence of exotic superfluid phases such as the Sarma \cite{Sarma1963} or FFLO phase \cite{Fulde1964,Larkin1965}. Eventually, for large enough chemical potential difference, superfluidity is completely destroyed. The stability of some exotic superfluid phases like FFLO is greatly enhanced by lowering the dimensionality of the gas \cite{Parish2007b,Koponen2008}.

The search for exotic superfluids motivates our study of spin-imbalanced atomic Fermi gases in two dimensions. In addition, the 2D case becomes particularly interesting in the case of strong interactions \cite{Conduit2008,He2008,Fischer2014,Sheehy2015}. In an atomic gas, Feshbach resonances enable tuning the interactions over a wide range and studying the effect of chemical potential imbalance beyond the described weak coupling BCS limit. Unlike the 1D case, exact solutions do not exist, and mean field models that do well in 3D fail in 2D due to the enhanced role of quantum fluctuations \cite{Yin2014,Strack2014}.

\begin{figure}
\includegraphics{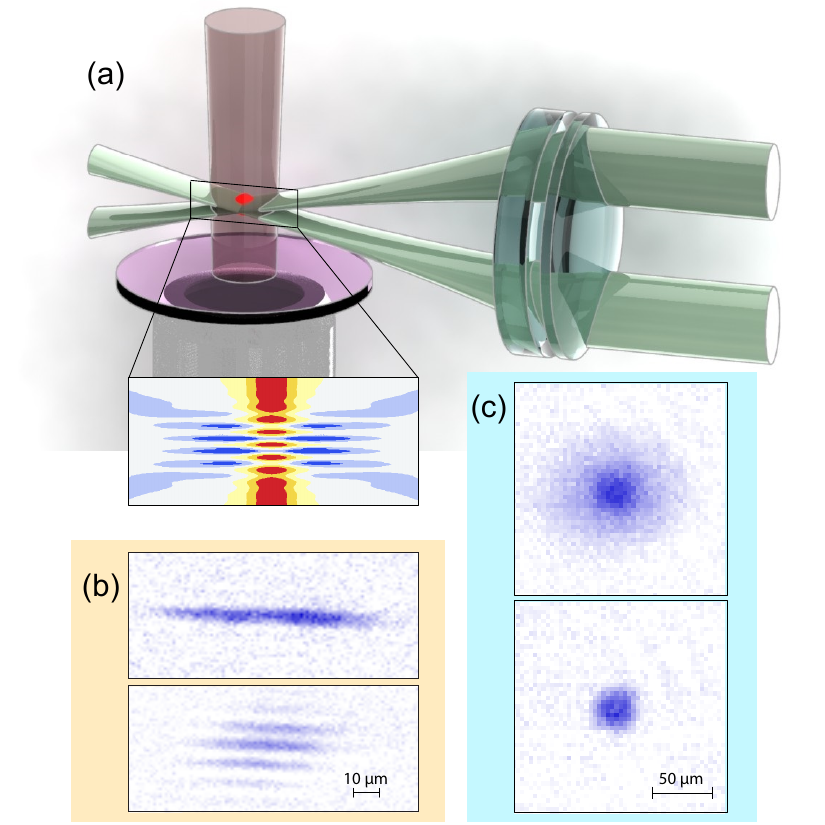}
\caption{Experimental setup. (a) A 1D optical lattice is formed at the intersection of two interfering blue-detuned beams (green), providing axial confinement. The lattice spacing can be dynamically tuned by changing the angle between the beams. The atoms are radially confined by a red-detuned beam (brown) in the vertical direction. A high-NA objective (grey) is used to image the in-plane density distribution. The inset shows a section of the optical potential with color scale from red (attractive) to blue (repulsive). (b) Side absorption images illustrating our capability to load and resolve single (above) and multiple pancakes (below) after adiabatically increasing the lattice spacing to $\sim 12~\mu$m. (c) In-situ absorption images of majority (above) and minority (below) clouds along the vertical direction at 755~G and polarization $P = 0.6$. \label{fig:schematic}}
\end{figure}

\begin{figure*}
\includegraphics{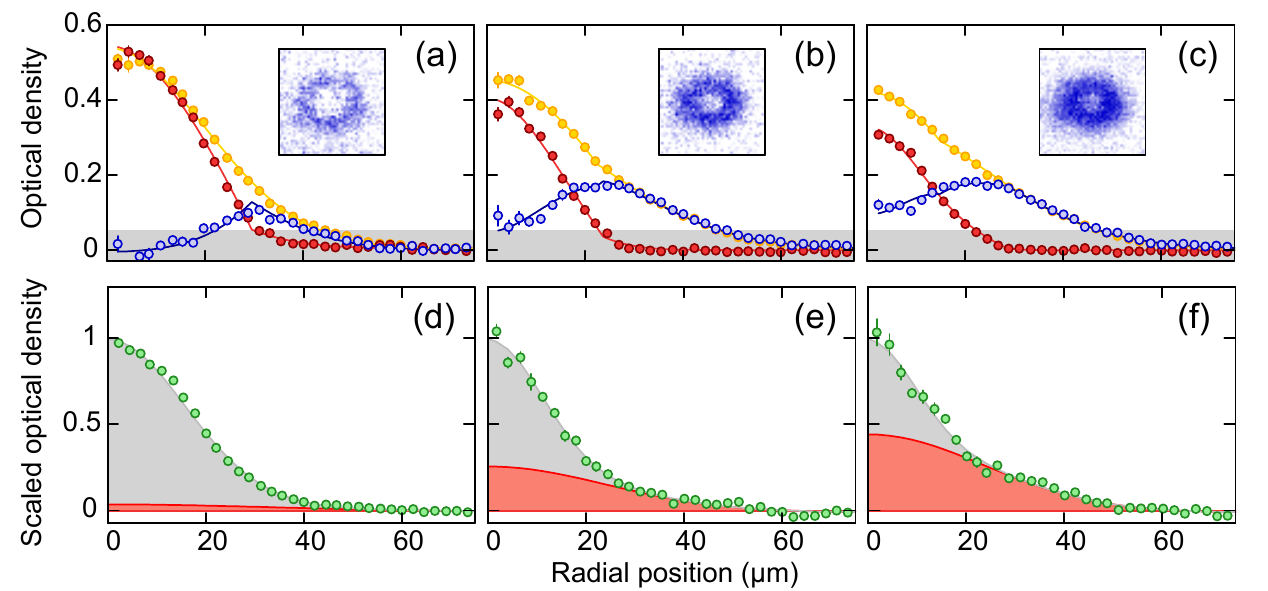}
\caption{Phase separation and condensation versus global polarization. (a)-(c) Azimuthal average of in-situ OD of majority (yellow) and minority (red) clouds, and of their difference (blue) for $P = 0.25$, 0.55, and 0.75 respectively taken at a field of 780~G, with fits to the sum of a Gaussian and a Thomas-Fermi profile. The insets show the corresponding two-dimensional OD difference. The gray shaded region represents the systematic error in the determination of OD differences. The radial position is measured along the minor axis of the elliptical contour lines used for azimuthal averaging. (d)-(f) OD of the minority cloud after 3~ms time of flight normalized to its peak value, with a double Gaussian fit to the data. The thermal component is shaded in red, while the condensate is shaded in gray. Error bars represent the standard deviation of the mean in evaluating the azimuthal average. All distributions represent an average of 30 experimental realizations.\label{fig:AzAv}}
\end{figure*}

Spin-imbalanced Fermi gases have been extensively studied both theoretically \cite{Bedaque2003,Carlson2005,Sheehy2006,Sheehy2007,Wu2016} and experimentally \cite{Zwerger2012}. Experiments in 3D have observed vortex lattices in spin-imbalanced superfluids \cite{Zwierlein2006} as well as phase separation between the superfluid and normal phases in the trapped gas \cite{Shin2006,Partridge2006pairing,Zwierlein2006direct}. Subsequent experiments quantitatively mapped out the  phase diagram of the 3D gas \cite{Shin2008,Olsen2015} and measured the equation of state of the imbalanced gas \cite{Navon2010,Nascimbene2010}. In 1D, phase separation was also observed, displaying an inverted phase profile in the trap compared to 3D \cite{Liao2010}. Recent experiments have started to explore 2D Fermi gases \cite{Martiyanov2010,Frohlich2011,Feld2011, Dyke2011,Sommer2012,Zhang2012,Makhalov2014,Boettcher2016,Fenech2016,Dyke2016}, mostly focusing on the spin-balanced case, where pair condensation has been observed \cite{Ries2015} and the BKT nature of the transition to the superfluid state was explored \cite{Murthy2015,Wu2015}. The properties of the polaron were characterized in experiments studying the extreme imbalance limit \cite{Koschorreck2012}. A non-interacting polaron model was found to be adequate for describing high-polarization 2D Fermi gases in the BCS regime and a spin-balanced central core was observed on the BEC side \cite{Ong2015}.


Our experiments measure the density profile of a single-layer 2D gas, revealing a spin-balanced core at low to intermediate polarizations. We explore the stability of this spin-balanced core for varying chemical potential, chemical potential difference between the spin species, and interaction strengths across the Feshbach resonance. On the BEC side of the 3D unitarity point, measurement of the momentum distribution in time-of-flight reveals a pair condensate. Condensation is observed past the disappearance of phase separation, implying that unpaired majority atoms become dissolved in the condensate, forming a polarized condensate.


We realize a strongly-interacting Fermi gas using a mixture of the lowest two hyperfine ground states of $^6$Li, $\left| 1\right>$ and $\left|2\right>$. The global spin imbalance, $P$, is defined as $P=\frac{N_{\uparrow}-N_{\downarrow}}{N_{\uparrow}+N_{\downarrow}}$, where $N_{\uparrow}$ is the population in the majority state ($\left| 1\right>$), and $N_{\downarrow}$ is the population in the minority state ($\left| 2\right>$). $P$ can be varied continuously from a balanced to an almost completely polarized gas. The interaction strength is varied by tuning the $s$-wave scattering length using a broad Feshbach resonance centered at 832~G.

To create the ultracold sample, we load atoms from a magneto-optical trap into a 1~mK deep crossed optical dipole trap. Starting from a spin-balanced mixture, the spin populations are imbalanced by transferring a variable fraction of the $\left\lvert 2\right>$ atoms into a third hyperfine state, $\left\lvert3\right>$, employing a diabatic Landau-Zener sweep \cite{supp}. The $\left\lvert3\right>$ atoms are then removed from the trap with a resonant light pulse before proceeding with all-optical evaporation at the Feshbach resonance.
The imbalanced mixture is transferred to a highly anisotropic optical trap with aspect ratio $\omega_x:\omega_y:\omega_z=1:3:30$. The large confinement anisotropy allows efficient transfer into a single well of a 1D optical lattice with a 12~\textmu m lattice spacing formed by two 532~nm laser  beams intersecting at a shallow angle that can be dynamically adjusted to change the lattice spacing [Fig.~\ref{fig:schematic}(a)]. In the plane, the atoms are confined by a vertical 1070~nm beam with a 100~\textmu m waist. Subsequently, the Feshbach field is adjusted to set the interaction strength in the gas and the lattice spacing is decreased to 3.5~\textmu m, resulting in trapping frequencies ($\omega_x, \omega_y, \omega_z) = 2\pi\cdot(124~\textrm{Hz}, 147~\textrm{Hz}, 22.5~\textrm{kHz})$. To ensure that the gas is in the 2D regime the chemical potential of the majority atoms, $\mu_{0\uparrow}$, is kept below the axial vibrational level spacing $\hbar\omega_z$. This condition is satisfied by keeping the majority atom number fixed to $\sim 9\times10^3$, resulting in $\mu_{0\uparrow}/\hbar\omega_z < 0.7$ over the full parameter regime. Our ability to load a single layer is confirmed by taking an absorption image of the cloud on-edge through an auxiliary imaging system. Example images of single and multiple loaded layers are shown in Fig. \ref{fig:schematic}(b).

We take absorption images of the minority and majority density distributions using two consecutive resonant pulses to obtain the optical density (OD) of the sample [Fig.~\ref{fig:schematic}(c) and \cite{supp}], which is directly proportional to the two-dimensional atomic density. The minority image is always taken first, although we have checked that the effect of heating due to the first imaging pulse is not measurable within our experimental noise. 


\begin{figure}
\includegraphics{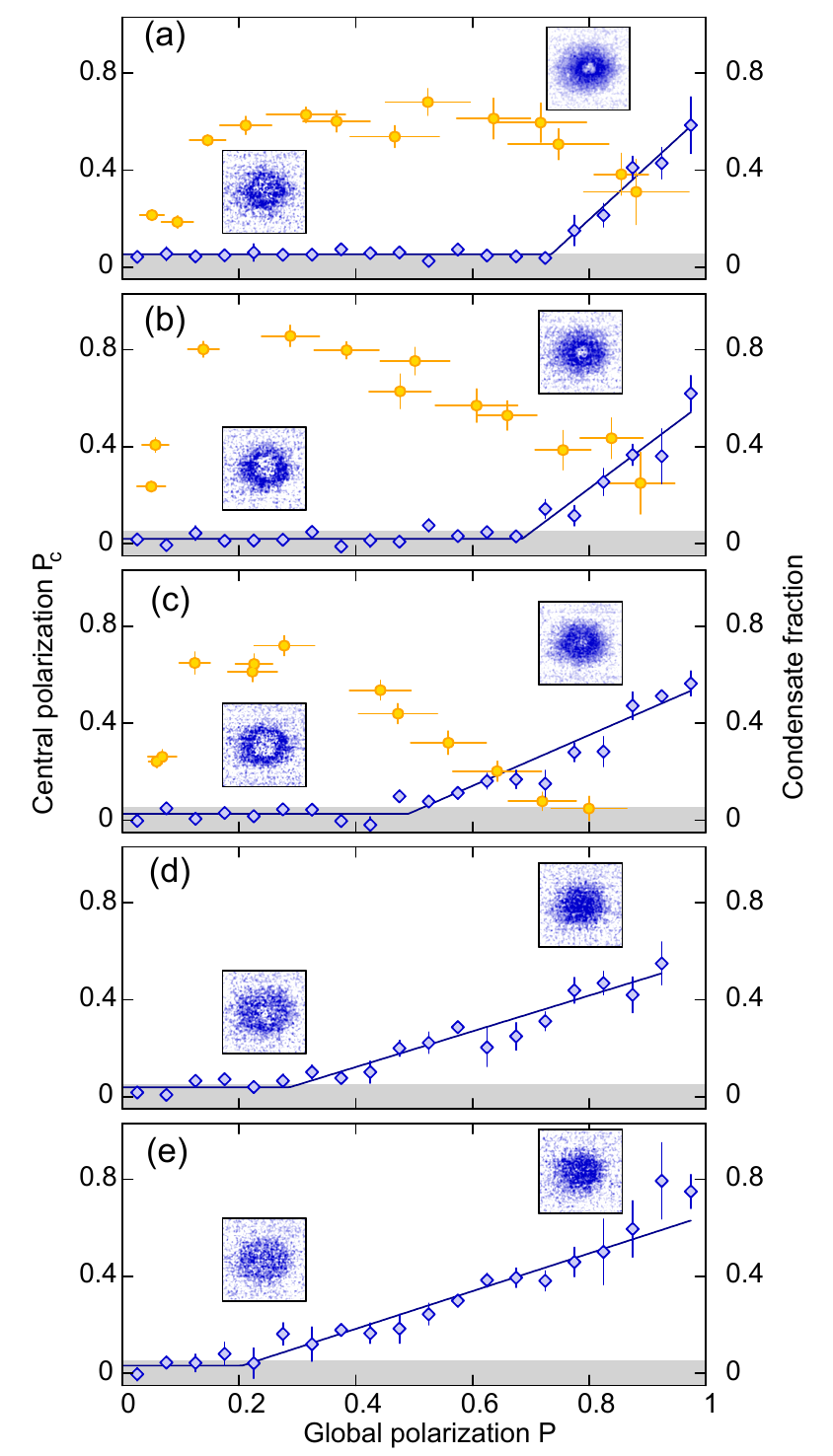}
\caption{Central polarization and condensate fraction vs. global polarization. Central polarization (blue diamonds) and condensate fraction (yellow circles) are shown for the Feshbach fields of (a) 730~G, (b) 755~G, (c) 780~G, (d) 830~G and (e) 920~G. The points represent the average of 5-15 experimental realizations, and error bars are the standard deviation of the mean. The gray region indicates the experimental uncertainty in the determination of the central polarization. The blue line is a bilinear fit to the data to determine $P_\text{c}$. For condensate fraction data, each point is obtained from a bootstrap analysis of 30 experimental shots. Error bars represent the standard deviation of the bootstrap distribution. The insets show the OD difference at $P=0.25$ and $0.75$. \label{fig:CvsG}}
\end{figure}


An in-situ image reveals the density of each spin component $n_{\uparrow}(r)$ and $n_{\downarrow}(r)$ and the local polarization $p(r)=\frac{n_{\uparrow}-n_{\downarrow}}{n_{\uparrow}+n_{\downarrow}}$ after azimuthally averaging over elliptical contour lines. Fig.~\ref{fig:AzAv}(a)-(c) shows in-situ density profiles for three different polarizations at 780~G. For all shown polarizations we observe a dip in the center of the difference OD. For $P = 0.25$ the central polarization is consistent with zero, while for $P = 0.75$ we observe a clear difference in central density for minority and majority components. These spatially varying profiles can be understood in the local density approximation. While the difference between the chemical potentials of the two species remains fixed throughout the trap, the average chemical potential is scanned by the trap. Thus one can expect shells of coexisting phases in the trap. The insets in Fig.~\ref{fig:AzAv} show an example of such structure where a balanced phase exists in the trap center, surrounded by a partially polarized gas which is, in turn, enclosed by a shell of fully polarized gas of majority atoms. A Fermi-Dirac fit to the tail of the radial majority density profile yields $T/T_F = k_BT/\mu_{0\uparrow}=0.18(5)$ independent of polarization, and only weakly dependent on the Feshbach field \cite{supp}. We note that this definition of $T_F$ deviates from the definition via the central density ($T_F^0$) used elsewhere \cite{Ries2015}. For our balanced data on the BEC side, we get $T/T_F^0 =$ 0.10(3)  \cite{supp}.

The existence of a spin-balanced core strongly suggests the presence of a condensate in that region of the trap. To probe pair condensation more directly, we measured the density of the minority component after a 3 ms time-of-flight. Unlike Ref. \cite{Ries2015}, we did not perform a rapid ramp to the BEC side, but simply released the gas from the trap. The expansion along the axial direction of the 2D gas leads to a rapid reduction of the density of the gas during time of flight, and the pair center of mass momentum distribution is not significantly affected by scattering events. We observe bimodal distributions that fit well to the sum of two Gaussian profiles. Examples are shown in Fig.~\ref{fig:AzAv}(d)-(f) corresponding to the same parameters as the in-situ images. We find a narrow condensed mode whose size remains roughly constant as the time of flight is increased and a wider thermal component that expands rapidly \cite{supp}. This allows us to define a condensed fraction as the ratio of minority atoms in the condensate mode to the total number of minority atoms. The paired nature of the condensate is confirmed by the observation that both the optical density and the width of the narrow mode match between majority and minority clouds.



\begin{figure}
\includegraphics{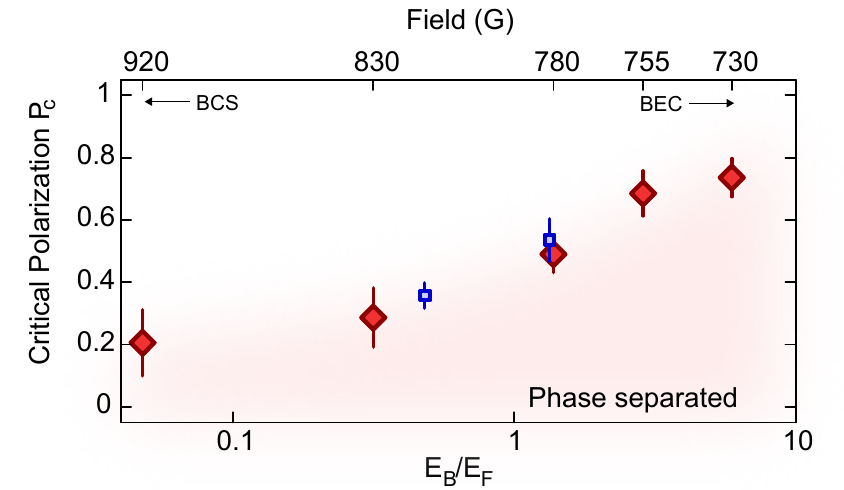}
\caption{Critical polarization ($P_\text{c}$) for phase separation vs. $E_B/E_F$ (log-scale). The global polarization at which phase separation ends is extracted from a bilinear fit to the central polarization data in Fig.~\ref{fig:CvsG} (red diamonds). Corresponding magnetic fields for these points are shown on the secondary $x$-axis. For comparison, we extracted two points from the data of ref. \cite{Ong2015} (blue squares). The error bars are given by the fit uncertainty. \label{fig:CPreduced}}
\end{figure}

We have studied the stability of the spin-balanced condensate to chemical potential imbalance across the BEC-BCS crossover. The chemical potential imbalance is scanned by changing the minority atom number and hence the global polarization $P$. The tight confinement of the gas along the axial direction allows for a two-body bound state with binding energy $E_B$ even above the Feshbach resonance, unlike the 3D case. The absence of a unitarity point in the quasi-2D case makes the distinction between the BEC and BCS regimes more \textit{ad hoc} than in 3D. We choose to characterize the interaction strength using the ratio $E_B/E_F$ , where $E_F = \hbar \sqrt{2\omega_x \omega_y N_{\uparrow}}$ is the Fermi energy of the majority atoms in the non-interacting gas. We identify the BEC regime with $E_B/E_F\gg1$ and the BCS regime with $E_B/E_F\ll 1$. The central polarization of the gas $p(0)$ is shown in Fig.~\ref{fig:CvsG}  vs. $P$ for Feshbach fields of 730~G, 755~G, 780~G, 830~G, and 920~G. The respective values of $E_B/E_F$ are $\sim$ 6(1), 2.9(7), 1.4(3), 0.32(7) and 0.05(2).  We find that $p(0)$ is consistent with zero within experimental uncertainty for a range of $P$ less than a field-dependent critical polarization $P_\text{c}$. In the BEC regime, the gas may be thought of as an interacting Bose-Fermi mixture of deeply bound dimers and excess majority atoms, with strong atom-dimer repulsion leading to the observed profiles. This picture is supported by comparison of a mean field model with the data in the BEC regime \cite{supp}. In the BCS regime, the superfluid gap prevents fermionic quasiparticles from entering the superfluid below the Clogston limit. We find that $P_\text{c}$ decreases as the BCS limit is approached as summarized in Fig.~\ref{fig:CPreduced}. Our observed critical polarization is consistent with a previous measurement~\cite{Ong2015} for comparable values of $E_B/E_F$.


We observe pair condensation that persists to high values of $P$ at $730$~G, $755$~G and $780$~G [Fig.~\ref{fig:CvsG}(a)-(c)], even beyond $P_\text{c}$, pointing to a polarized condensate. For $B=830$~G and $920$~G, no bimodality is observed. This can be anticipated for expansion in 3D since there is no bound state beyond the Feshbach resonance and the fragile dimers that exist in the trapped system break after release. The measured condensate fraction for a balanced gas is compatible with the fraction that has been measured recently \cite{Ries2015} for comparable $T/T_F^0$. Similar to experiments in 3D \cite{Zwierlein2006}, we find that the condensate fraction does not drop monotonically with increasing $P$ as one would expect naively, but rather peaks at a non-zero $P$. The harmonic confinement of the clouds may explain this observation. Although the absolute temperatures we measure are independent of $P$ \cite{supp}, increasing $P$ leads to a shrinking minority cloud whose wings experience a higher majority density, and therefore a higher local critical temperature. 

\begin{figure}
\includegraphics{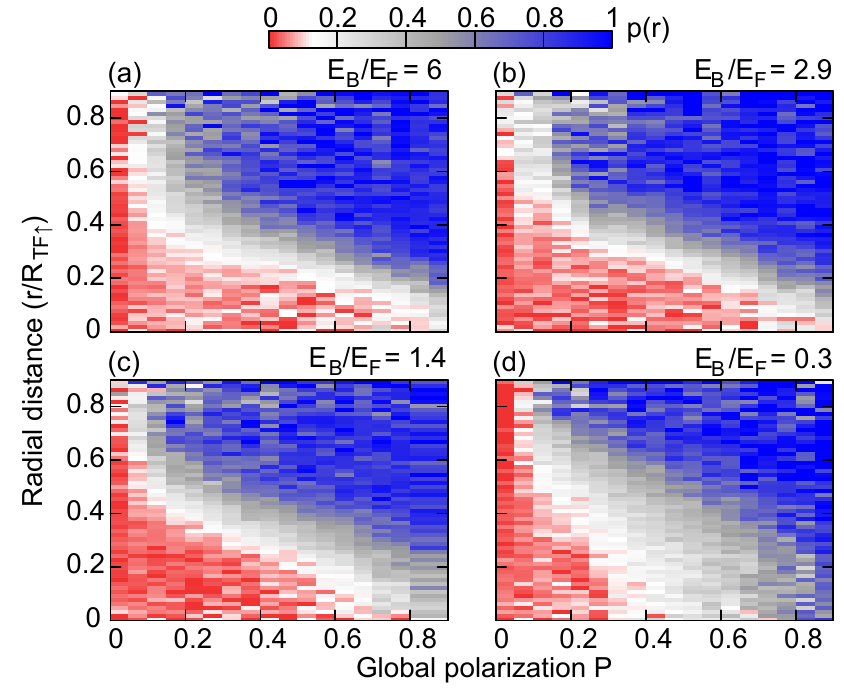}
\caption{Phase diagram of an imbalanced 2D Fermi gas for four different interactions determined by the Feshbach fields (a) 730~G, (b) 755~G, (c) 780~G and (d) 830~G. The corresponding interaction strengths $E_B/E_F$ are shown in the top-right corner of each panel. The color indicates the local polarization of the gas in the trap $p(r)$ as a function of the scaled position in the trap $r/R_{\text{TF}\uparrow}$ and the global polarization $P$. We distinguish three different phases: a balanced condensate (red), a partially polarized phase (white to gray) and a fully polarized normal gas (blue).\label{fig:PD}}
\end{figure}

Fig.~\ref{fig:PD} shows an experimental phase diagram of a spin-imbalanced 2D Fermi gas for four different values of the interaction strength $E_B/E_F$. These phase diagrams show the local polarization $p(r)$ as a function of the global polarization $P$  and the position in the trap $r$ scaled by the Thomas-Fermi radius of a fully polarized gas $R_{TF\uparrow}$, defined as $V(R_{TF\uparrow}) = \mu_{0\uparrow}$ \cite{supp}.  The global polarization $P$ is the experimental parameter that determines the chemical potential difference $h = (\mu_{\uparrow}-\mu_{\downarrow})/2$, while $r$ fixes the average local chemical potential $\mu = (\mu_{\uparrow}+\mu_{\downarrow})/2$, so we can interpret these diagrams as ``$\mu-h$" phase diagrams expressed in terms of experimentally measured quantities. The partially polarized phase occupying the part of the phase diagram between the balanced condensate and the fully polarized normal gas and depending on $E_B/E_F$ may be a Sarma phase induced by quantum or thermal fluctuations, an FFLO phase or a Fermi liquid phase.


Unlike the 3D case \cite{Shin2008}, we have not observed discontinuities in the polarization or density profiles. Zero temperature phase diagrams in refs. \cite{Conduit2008,He2008,Sheehy2015} predict a first order phase transition between the superfluid and normal phases driven by the change in the average local chemical potential in the trap. This would be manifested by a sudden jump of the local polarization from zero in the superfluid to a finite polarization in the normal phase. In 3D, the first order transition only occurs for temperatures below the tricritical point \cite{Parish2007}. If such a tricritical point exists in 2D, it is possible that the temperature of our clouds is not low enough to observe the first order transition or that the one-dimensional nature of the interface between the superfluid and normal phases makes it very susceptible to fluctuations that smear out discontinuities when averaging over trap contour lines. We also note that strong quantum fluctuations in 2D can in principle drive the superfluid to normal transition continuous \cite{Strack2014}.


In conclusion, we have observed pair condensation in an imbalanced Fermi gas across the BEC-BCS crossover, accompanied by phase separation in the trap. In future work, it will be interesting to study the effect of thermal fluctuations on superfluidity in the imbalanced 2D gas and to determine if imbalance affects the BKT nature of the transition. Another interesting direction is the investigation of the partially polarized gas between the balanced condensate and the fully polarized normal gas. The partially polarized gas may host a variety of phases, including Sarma or FFLO states, whose stability is enhanced in lower dimensions. 


\begin{acknowledgments}
We thank David Huse, P\"aivi T\"orm\"a and Philipp Strack for useful discussions and Martin Zwierlein for critical reading of the manuscript. We thank John Thomas, Ilya Arakelyan and coworkers for sharing their data. This work was supported by the NSF and the AFOSR Young Investigator Research Program. W.S.B. was supported by an Alfred P. Sloan Foundation fellowship and P.T.B. was supported by the DoD through the NDSEG Fellowship Program.
\end{acknowledgments}


\bibliography{../../bibliography}


\pagebreak
\clearpage
\setcounter{equation}{0}
\setcounter{figure}{0}
\renewcommand{\thefigure}{S\arabic{figure}}
\renewcommand{\theequation}{S\arabic{equation}}
\onecolumngrid
\begin{center}
\large{\textbf{Supplemental Material :  Phase separation and pair condensation in a spin-imbalanced 2D Fermi gas}}
\\

\end{center}

\twocolumngrid


\begin{figure*}
\includegraphics{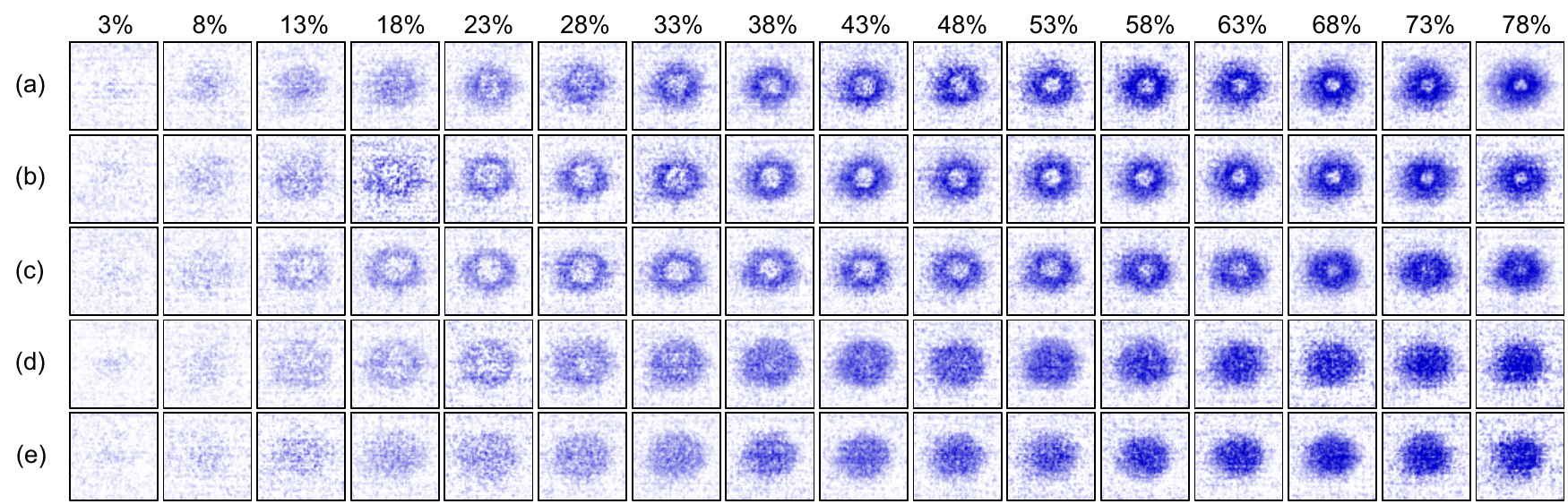}
\caption{Two dimensional OD difference images for varying global polarization (columns) and magnetic fields (rows). The magnetic fields are (a) 730~G, (b) 755~G, (c) 780~G, (d) 830~G and (e) 920~G.  The global polarization is labelled at the top of each column.\label{fig:figMatrix}}
\end{figure*}

\section{Preparation of a 2D Fermi gas}

We begin by loading $10^6$  $^6$Li atoms from a compressed magneto-optical trap into a 1~mK deep crossed optical dipole trap. The trap is created using two 90~W, 1070~nm counterpropagating laser beams with orthogonal polarizations intersecting at an angle of $11^\circ$. The beam waist at the atoms is 80~$\mu$m.  After optically pumping the atoms to the lowest two hyperfine states $\left|1\right\rangle$ and $\left|2\right\rangle$, we produce a balanced mixture of these states using ten consecutive diabatic Landau-Zener sweeps at 537~G, with a sweep rate chosen to approximately transfer 50\% of the atoms from one state to the other. We then imbalance the mixture by transferring a fraction of the atoms from state $\left|2\right\rangle$ into a third hyperfine state $\left|3\right\rangle$ using another diabatic Landau-Zener sweep. We subsequently remove the atoms in state $\left|3\right\rangle$ with a resonant pulse. We control the amount of imbalance using the sweep rate. The imbalanced mixture is then evaporatively cooled at 800~G close to degeneracy. We load the atoms into a highly anisotropic light sheet trap ($\omega_x:\omega_y:\omega_z=1:3:30$) generated using the same laser to compress the atoms axially in preparation for loading them into a single well of the optical lattice. To independently control the radial confinement, we ramp up a round, vertical laser beam with a $100~\mu$m waist also derived from the same laser. 

In order to produce a large, tunable axial confinement, we use an ``accordion" optical lattice using light at 532~nm [Fig. 1 of main text]. The spacing of the different layers of the accordion can be changed by controlling the angle at which the lattice beams intersect, allowing us to load a single layer reproducibly at large spacing (12~$\mu$m) and then adiabatically reduce the lattice spacing to 3.5~$\mu$m to reach a quasi-2D geometry. At this stage, we turn off the light sheet confinement and all of the axial confinement is provided by the lattice. 

Finally, we perform a second evaporation in the combined 2D trap by ramping down the radial confinement to its final value in 200~ms. The trapping frequencies of our combined trap are $\omega_x = (2\pi)~124$~Hz, $\omega_y = (2\pi)~147$~Hz and $\omega_z = (2\pi)~22.5$~kHz. The final number of atoms in state $\left\lvert1\right>$ is held constant at $\sim 9\times 10^3$, while the number in state $\left\lvert2\right>$ is varied. The magnetic field is then ramped to its final value in 50~ms to set the interactions in the 2D gas.

\begin{figure}
\includegraphics{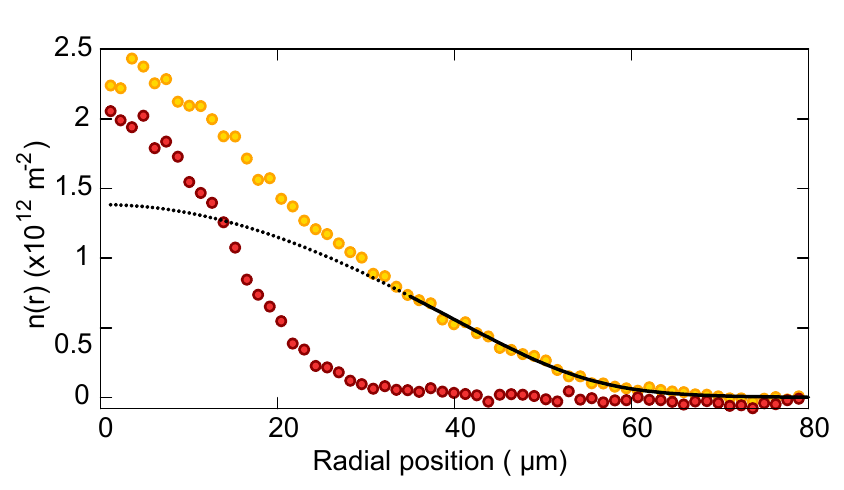}
\caption{Fermi-Dirac fit to the fully-polarized tail of a majority profile to obtain temperature. Data depicted for $P=0.48$ at 780~G. The majority density $n_{\uparrow}(r)$  is shown in yellow and the minority density $n_{\downarrow}(r)$ in red. The black solid line is the fit to the majority density using the expression in Eq.~(\ref{eq:Fermi}) beyond the Thomas-Fermi radius of the minority cloud. The black dotted line is the fit extrapolated to the trap center. \label{fig:TempFit}}
\end{figure}


\section{Comparison of condensate fraction}
To compare the measured condensate fraction for a balanced gas with ref. \cite{Ries2015}, we calculate $T/T_F^0$ as defined in this reference. This yields $T/T_F^0 = 0.10(3)$ for 730-780~G, where $T_F^0 = \frac{2\pi n_0 \hbar^2}{m k_B}$, with $n_0$ the central density of one spin component for a balanced gas, and $m$ the mass of $^6$Li.
For the calculated values of $\ln{(k_F a_{2D})} = -2$  $\ldots$  $-0.5$ for our data we find a condensate fraction of $0.25(5)$ consistent with ref. \cite{Ries2015} for the given temperature.

\section{Imaging}
We image the clouds at the final magnetic field using an objective with a numerical aperture of 0.5, corresponding to a resolution of $\sim800$~nm at 671~nm. We take images of both states consecutively in each experimental run. For this purpose, we take advantage of the particle imaging velocimetry mode of an Andor Neo 5.5 sCMOS camera. The two imaging pulses are obtained from the same beam by fast switching of the frequency by $\sim80$~MHz using an acousto-optic modulator. The imaging pulses are 10~$\mu$s long and are separated by 10~$\mu$s. To calibrate the optical depth of the gas to the absolute density, we use the known density of a band insulator in an optical lattice with 752~nm spacing.

\section{Fitting of in-situ profiles}
Fig. \ref{fig:figMatrix} displays in-situ OD difference profiles in two-dimensions as a function of interactions (rows) and global polarization (columns). We average each image azimuthally over the elliptical contour lines of the trap. The resulting profiles are fit to the sum of a one-dimensional Gaussian and an inverted parabola. The fits give us the atom number in each species. We define a global polarization $P$ in the trap using $P = (N_{\uparrow}-N_{\downarrow})/(N_{\uparrow}+N_{\downarrow})$. We post-select each in-situ profile to be averaged with other profiles within a range of $P = 0.05$. This leads to 20 global polarization bins, starting at $P=0.025$ up to $P = 0.975$, each containing 
$5-15$ images. From the radial profile of the averaged optical density, we define the local atomic density for each species $n_{\uparrow}(r) = 2\pi\text{OD}_{\uparrow}(r)/3\lambda_{0}^2$ and  $n_{\downarrow}(r) =  2\pi\text{OD}_{\downarrow}(r)/3\lambda_{0}^2$  (red and yellow points in Fig. 2(a)-(c) of main text), with $\lambda_{0}=$ 671~nm and the local polarization as $p(r) = (n_{\uparrow}(r)-n_{\downarrow}(r))/(n_{\uparrow}(r)+n_{\downarrow}(r))$.


\begin{figure}
\includegraphics{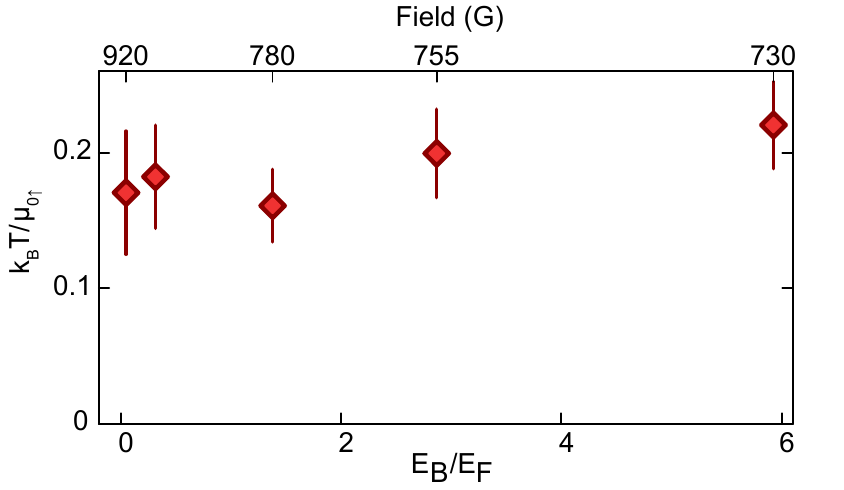}
\caption{Interaction dependence of $k_B T/\mu_{0\uparrow}$. The temperature ($T$) and the majority chemical potential ($\mu_{0\uparrow}$) are obtained from a Fermi-Dirac fit to the fully-polarized region of the cloud. \label{fig:ToverMu}}
\end{figure}

\begin{figure}
\includegraphics{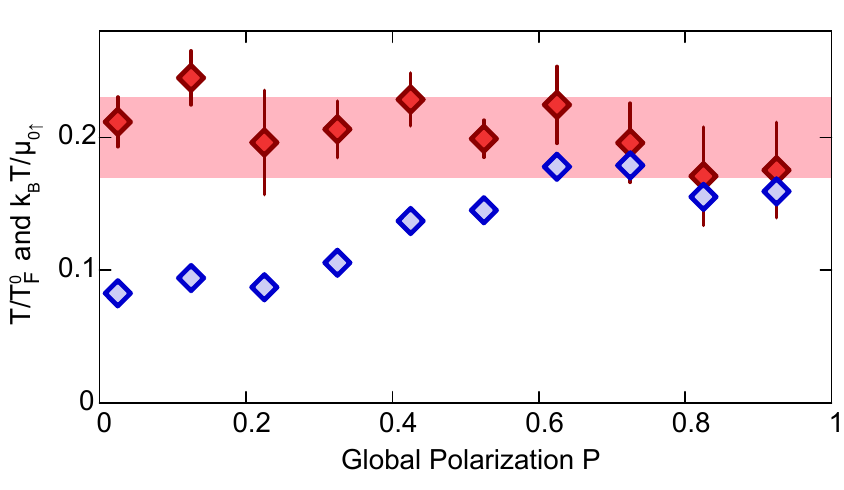}
\caption{Polarization dependence of $k_B T/\mu_{0\uparrow}$ (red diamonds) and $T/T_F^0$ (blue diamonds) for the 755~G dataset. $k_B T/\mu_{0\uparrow}$ stays essentially constant over the whole polarization range while $T/T_F^0$ drops towards the balanced regime due to increased density in the center of the cloud. \label{fig:ToverP}}
\end{figure}

\section{Measuring temperature}
Using the in-situ profiles, we obtain the temperature of the gas along with the majority chemical potential ($\mu_{0\uparrow}$) by fitting the tail of an imbalanced gas [Fig. \ref{fig:TempFit}]. This works best when there is a sufficiently large fully-polarized region in the cloud ($P>0.4$), but we see no significant temperature change towards more balanced gases [Fig. \ref{fig:ToverP}]. The expression for the density of a non-interacting 2D Fermi gas in a trap at finite temperature is

\begin{equation} \label{eq:Fermi}
n_{\uparrow}(r) =  \frac{2\pi m}{h^2 \beta}\text{log}\left[1+\text{exp}\left(\beta\left(\mu_{0\uparrow}-V(r)\right)\right)\right]
\end{equation}
where $\beta = 1/k_BT$, $V(r)$ is the Gaussian trapping potential and $\mu_{0\uparrow}$ is the majority chemical potential at the center of the trap. Using small amplitude sloshing measurements and high resolution imaging of our trapping beam we determined the potential to be well described by $V(r) = \frac{1}{2}m\omega^2 \frac{w_0^2}{2} \left[ 1 - \exp\left(-2 (r/w_0)^2\right)\right]$ with $\omega=2\pi\times 147 \text{ Hz}$ and $w_0 = 101\text{ \textmu m}$. We find no significant dependence of $T$ or $\mu_{0\uparrow}$ on the polarization within the uncertainty of the fit and observe a weak dependence of $T$ on the interaction strength [Fig. \ref{fig:ToverMu}].

\begin{figure}
\includegraphics{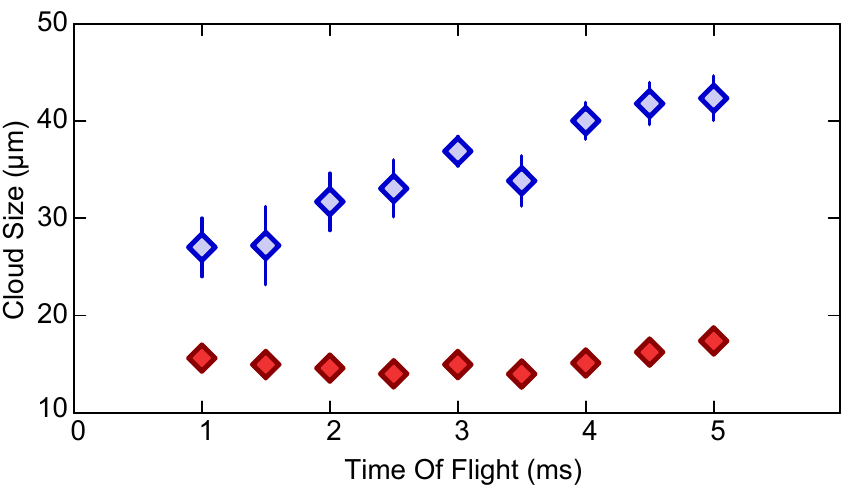}
\caption{Minority cloud size versus time of flight. Data displayed for 780~G and $P=0.3$. The time of flight data is fit to the sum of two Gaussians. Blue diamonds are the Gaussian width of the thermal part while the red diamonds are the Gaussian width of the condensate. The thermal cloud expands much more rapidly than the central mode during time of flight. \label{fig:sigmas}}
\end{figure}
\section{Measuring the condensed fraction}
We measure momentum space profiles by releasing the cloud from the trap and allowing a time of flight of 3~ms. Due to the tight confinement in the axial direction, the cloud expands fast along that axis and the interactions are effectively quenched. For the data at 730~G, 755~G and 780~G, we see clear signatures of bimodality in the minority density after time of flight. We quantify this bimodality by fitting the azimuthal average of the minority density after time of flight to a sum of two Gaussians. The narrow Gaussian fits the slowly expanding condensate while the broader Gaussian fits the rapidly expanding thermal cloud [Fig. 2(d)-(f) of main text]. The standard deviation of the two Gaussian modes are shown in Fig. \ref{fig:sigmas}. From the size of the thermal cloud, we extract a kinetic energy per particle of $\sim70$~nK. The condensed fraction is the ratio of the number of atoms in the condensate to the total atom number.

\section{Mean field model in the BEC regime}

\begin{table}
\begin{ruledtabular}
\begin{tabular}{lrrr}
Feshbach field & 730~G & 755~G & 780~G \\\hline
$\mu_\text{f}$ in kHz & 14.5(2) & 14.5(2) & 14.0(2) \\
$\tilde{g}_\text{bb}$ & 2.54(3) & 2.46(4) & 2.78(4) \\
$\tilde{g}_\text{bf}$ & 2.71(5) & 2.69(5) & 2.79(6) \\
$a_\text{bb}$ in $a_0$ & 3710(100) & 3580(110) & 4050(120) \\
$a_\text{bf}$ in $a_0$ & 3220(110) & 3200(120) & 3320(140) \\
$a_\text{bb}$ in $a_\text{ff}$ & 1.46(4) & 0.93(3) & 0.63(2) \\
$a_\text{bf}$ in $a_\text{ff}$ & 1.27(4) & 0.83(3) & 0.52(2) \\
\end{tabular}
\end{ruledtabular}
\caption{Mean field theory fitting results. The scattering length is comparable to the axial harmonic oscillator length for all magnetic fields. See text for definition of parameters.\label{tab}}
\end{table}

\begin{figure}
\includegraphics{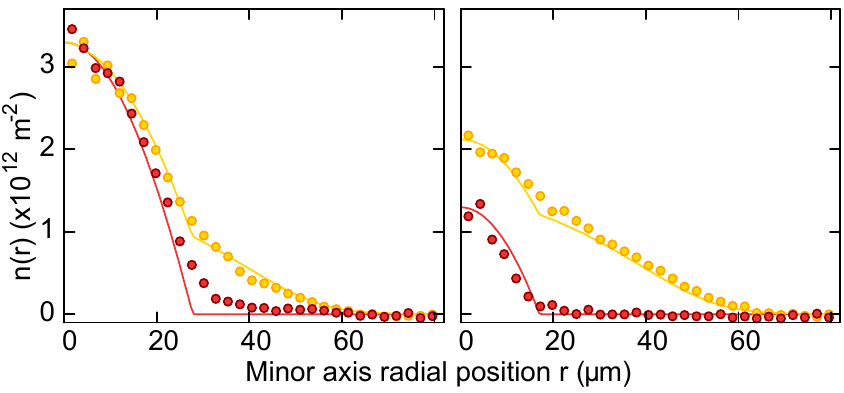}
\caption{Example mean field theory fits. Data displayed for 755~G and $P=0.2$ (left) and $P=0.8$ (right) with interaction parameters given in table~\ref{tab}. The majority ($n_f + n_b$) is shown in yellow and the minority ($n_b$) in red, points and line are experiment and theory, respectively. \label{fig:mean_field}}
\end{figure}

In the BEC regime it is possible to employ a simple zero-temperature mean field model \cite{Buechler2004,Shin2008b}. Here we limit our focus on a 2D Fermi gas with effective density-independent 3D scattering, that is approximately valid for 730~G and 755~G and describes our data well for these datasets.
At 780~G the scattering length becomes larger than the harmonic oscillator length in z-direction and it would be required to take into account corrections to the scattering.
The mean field model describes the gas as a mixture of bosons with mass $m_\text{b} = 2 m$ and fermions with mass $m_\text{f} = m$ with the assumption that all minority atoms are paired. Under the minimization condition of the total Gibbs free energy function of fermion density ($n_f$) and boson density ($n_b$), one obtains the following coupled linear equations for the chemical potentials. 

\begin{align}
\mu_\text{f0} &= \frac{2\pi\hbar^2}{m_\text{f}} n_\text{f} + g_\text{bf} n_\text{b} + V_\text{f}(r)\\
\mu_\text{b0} &= g_\text{bb} n_\text{b} + g_\text{bf} n_\text{f} +V_\text{b}(r)
\end{align}

with $g_\text{bf} = \frac{\sqrt{8\pi}\hbar^2 a_\text{bf}}{m_\text{bf} l_\text{z,bf}} = \frac{\hbar^2}{m} \tilde{g}_\text{bf} $,
$g_\text{bb}  = \frac{\sqrt{8\pi}\hbar^2 a_\text{bb}}{m_\text{b}  l_\text{z,b}} = \frac{\hbar^2}{m} \tilde{g}_\text{bb} $,
with $m_\text{bf} = \frac{2 m_\text{b} m_\text{f}}{m_\text{b} + m_\text{f}}$ and $  l_\text{z,b} = \sqrt{\frac{\hbar}{m_\text{b} \omega_z}} = 3650 a_0$ and 
$  l_\text{z,bf} = \sqrt{\frac{\hbar}{m_\text{bf}  \omega_z}} = 4470 a_0$ the effective harmonic oscillator lengths and $a_0$ is the Bohr radius. 
$V_\text{b,f}$ is the effective Gaussian trapping potential for bosons and fermions, respectively.

These equations can be solved analytically for all $r$ independently under local density approximation and directly give the radial density profiles of minority and majority spin component as a function of the global chemical potentials $\mu_\text{f0}$ and $\mu_\text{b0}$. 
We note that this model shows perfectly matching minority and majority densities in the center of the trap as well as partially polarized regions in the trap over a wide parameter range [Fig. \ref{fig:mean_field}].

We observe deviations from the predicted boson-boson and boson-fermion scattering lengths which were calculated for 3D to be $0.6 a_\text{ff}$ and $1.18 a_\text{ff}$ for boson-boson and boson-fermion interactions respectively \cite{Skorniakov1957, Petrov2003,Petrov2004}. These may be because we are not fully in the 3D scattering regime. Furthermore, Ref \cite{Shin2008b} which studied phase separation in 3D imbalanced clouds found that beyond mean field corrections can be significant. We find reasonable agreement with our data at 730~G, 755~G and 780~G when using simultaneous fitting of all data for each Feshbach field to determine consistent effective scattering parameters ($a_\text{ff}$ from \cite{Zuern2013}). Here the only polarization-dependent fit parameter is $\mu_\text{b0}$. In the model function we replace the non-interacting Fermi tail by the exact finite-temperature solution.

\end{document}